# Thermopower of Ce$_x$R$_{1-x}$B$_6$(R = La, Pr and Nd)

Moo-Sung KIM, $^+$Yuki NAKAI, $^+$Hideki TOU, $^+$Masafumi SERA, $^+$Fumiroshi IGA, $^+$Toshiro TAKABATAKE
and $^*$Satoru KUNII

*Dept. of Physics, University of Michigan, Ann Arbor, Michigan 48109, USA*
$^+$*Dept. of Quantum Matter, ADSM, Hiroshima University, Higashi-Hiroshima 739-8530, Japan*
$^*$*Dept. of Physics, Tohoku University, Sendai 980-8678, Japan*

The thermopower, $S$, of Ce$_x$R$_{1-x}$B$_6$ (R = La, Pr, Nd) was investigated. $S$ with a positive sign shows a typical behavior observed in the Ce Kondo system, an increase with decreasing temperature at high temperatures and a maximum at low temperatures. The $S$ values of all the systems at high temperatures are roughly linearly dependent on the Ce concentration, indicating the conservation of the single-impurity character of the Kondo effect in a wide $x$ range. However, the maximum value of $S$, $S_{\max}$, and the temperature, $T_{\max}$, at which $S_{\max}$ is observed exhibit different $x$ dependences between Ce$_x$La$_{1-x}$B$_6$ and Ce$_x$R$_{1-x}$B$_6$ (R = Pr, Nd). In Ce$_x$La$_{1-x}$B$_6$, $T_{\max}$, which is $\sim$ 8 K in CeB$_6$, decreases with decreasing $x$ and converges to $\sim$ 1 K in a very dilute alloy and $S_{\max}$ shows an increase below $x \sim 0.1$ after decreasing with decreasing $x$. In Ce$_x$R$_{1-x}$B$_6$ (R = Pr, Nd), $T_{\max}$ shows a weak $x$ dependence but $S_{\max}$ shows a roughly linear decrease in $x$. These results are discussed from the standpoint of the chemical pressure effect and the Ce-Ce interaction. $S$ in the long-range ordered phase shows very different behaviors between Ce$_x$Pr$_{1-x}$B$_6$ and Ce$_x$Nd$_{1-x}$B$_6$.

## 1. Introduction

For more than two decades, much attention has been given in studying the coexistence and competition between the single-site Kondo effect and the intersite exchange interaction.[1,2] Although the single-impurity Kondo problem is well understood by now, the Kondo lattice problem still remains to be clarified. The ground state of the Kondo lattice system is determined by the competition between two energy scales, the Kondo temperature, $T_K$, and the intersite interaction between the Kondo impurities. Experimental studies for Ce-La and other systems have been performed extensively to clarify how the ground state is varied from the single site to the lattice system. However, its mechanism still remains to be clarified. In this respect, the two-impurity Kondo problem which concerns the RKKY interaction between two Kondo impurities is one of the important themes in the field of heavy-electron physics.[3-6]

In the Kondo systems, the electrical resistivity, $\rho$, has been studied in detail, both experimentally and theoretically. The thermopower, $S$, has also been studied by many groups. In many of the Kondo systems, $S$ exhibits an increase with decreasing temperature, a maximum around $T_K$ and then a decrease down to lower temperature. When $T_K$ is rather high, a sign change is often observed at low temperatures. Although several mechanisms have been proposed,[7-11] the understanding of the above characteristics of $S$ is still controversial. The effect of the crystalline electric field (CEF) on $S$ has also been studied.[12-15] A shoulder or maximum due to the CEF effect often masks the characteristics of the ground state. In order to clarify the thermopower of the Kondo system, it is important to avoid the ambiguity from the CEF effect. In this sense, Ce$_x$La$_{1-x}$B$_6$ is a good material because the CEF $\Gamma_8$ quartet ground state is well separated from the excited $\Gamma_7$ doublet state and $T_K$ is low.[16] When $T_K$ is comparable to the intersite interaction between the Kondo impurities, a rich variety of interesting ground states appear. CeB$_6$ is one of the typical examples with such an interesting ground state. CeB$_6$ is famous for the existence of the unusual antiferro-quadrupolar (AFQ) ordered phase II dominated by the AF octupolar (AFO) and AF exchange interactions.[17-19] For the last several decades, how the character of the Kondo effect is varied by the substitution of La ions in Ce$_x$La$_{1-x}$B$_6$ has been studied.[20-22] For Ce$_x$La$_{1-x}$B$_6$, the $\ln T$ dependence of $\rho$ at high temperatures is independent of Ce concentration, indicating that the single-impurity Kondo effect is conserved at high temperatures. Recently, differentiating from the Kondo effect in La-doped Ce compounds, we have studied how the Kondo effect is affected in the localized magnetic system using Ce$_x$R$_{1-x}$B$_6$(R = Pr, Nd).[23-25] PrB$_6$ exhibits two successive phase transitions at $T_{Cm}$ = 4.2 K and $T_{IC}$ = 7 K.[26] The commensurate (C) magnetic ordered state is realized below $T_{Cm}$ and the incommensurate (IC) one below $T_{IC}$. It is suggested that the $O_{xy}$-type AFQ interaction plays an important role in the C phase from the similarity of the noncollinear magnetic structure with an easy axis along the twofold one to that in phase III of CeB$_6$.[26,27] Recently, we reported the results for Ce$_{0.7}$Pr$_{0.3}$B$_6$.[23] The AFQ and magnetic ordered phases coexist and $\rho$ exhibits the destruction of the Kondo singlet state in the magnetic long-range-ordered (LRO) phase in this compound. These indicate the strong Ce-Pr coupling in Ce$_{0.7}$Pr$_{0.3}$B$_6$. NdB$_6$ exhibits a type-I AF magnetic structure below $T_N$ = 8 K where the $O_2^0$-type FQ interaction plays an important role.[28,29] Recently, we reported the following results of the $x$-dependent physical properties of Ce$_x$Nd$_{1-x}$B$_6$.[24,25] $\rho$ per Ce ion is nearly independent of Ce concentration at high temperatures as was observed in $Ce_xLa_{1-x}B_6$, indicating that the Kondo effect at high temperatures has a single-impurity character. Namely, the Kondo effect at high temperatures is hardly affected by the surrounding Nd ions. The magnetic phase diagram of $Ce_xNd_{1-x}B_6$ shows that the AFQ ordering does not coexist with the magnetic ordering for $x < 0.8$, which differs from the case of $Ce_{0.7}Pr_{0.3}B_6$.[23] For $x \leq 0.88$, two successive magnetic phase transitions appear at $T_N^H$ and $T_N^L$.[24,25] For $x \leq 0.5$, the LRO phase at the higher transition temperature $T_N^H$ is mainly induced by Nd ions and the Kondo singlet state survives below $T_N^H$. However, at the lower transition temperature $T_N^L$, the Ce ions themselves contribute to the formation of the magnetic order and the Kondo singlet state is destroyed below $T_N^L$.

We have two purposes in the present work. The first is to clarify the mechanism of the competition between the Kondo effect and the intersite interaction between the Ce ions and how the ground state changes from the dilute Kondo system to the lattice system in $Ce_xLa_{1-x}B_6$. This system is one of the best for clarifying the $S$ of the Kondo system both experimentally and theoretically because $CeB_6$ has a well-separated CEF $\Gamma_8$ ground state with low Kondo temperature. The second is to clarify how $S$ is affected by the $R(=Pr, Nd)$ doping in $CeB_6$ where Ce ions are surrounded by the magnetic $R$ ions.

## 2. Experimental Details

Most of the samples used in the present experiments are the single-crystalline but several polycrystalline samples, namely, $Ce_xLa_{1-x}B_6$ ($x = 0.01, 0.03, 0.2, 0.9$) and $Ce_xPr_{1-x}B_6$ ($x = 0.03, 0.1$), are also used. The details of the preparation of the single-crystalline $Ce_xR_{1-x}B_6$ ($R$ =La, Nd and Pr) were described in a previous report.[30]

The thermopower measurement was performed from 1.5 to 300 K by a usual differential method in zero magnetic field. The sample was suspended between two copper blocks across which $1 \sim 5\%$ of the measurement temperature was set in a temperature gradient. The thermoelectric voltage was measured through gold wires fixed on the sample. The absolute thermopower of the gold wires used in the present experiment was caliblated against pure lead using the thermopower data of Roberts.[31]

## 3. Experimental Results

### 3.1 Thermopower of $Ce_xR_{1-x}B_6(R = La, Pr, Nd)$ in paramagnetic region

Figure 1 shows the temperature dependence of the thermopower, $S$, of $Ce_xR_{1-x}B_6(R = La, Pr$ and Nd) in the form of $S$ vs $\ln T$.

Figure 2 shows the $x$ dependences of $S_{max}$, $T_{max}$, $S(20 K)$ and $S(50 K)$ of $Ce_xR_{1-x}B_6(R = La, Pr, Nd)$. Here, $S_{max}$ is the maximum value of $S$. $T_{max}$ is the temperature at which $S$ takes a maximum. $S(20 K)$ and $S(50 K)$ are the $S$ values at $T = 20$ K and 50 K, respectively.

In $CeB_6$, $S$ with a positive sign exhibits an increase with decreasing temperature and a large peak of $\sim 60$ $\mu V/K$ at $T \sim 8$ K, indicating the existence of the Kondo effect. With decreasing temperature, two anomalies are observed at $T_Q = 3.3$ K and $T_N = 2.3$ K and a change of sign is observed below $T_N$, which are in good agreement with previous reports.[32–34] It is noted that a tiny peak is observed at $T_N$ which was not reported before.

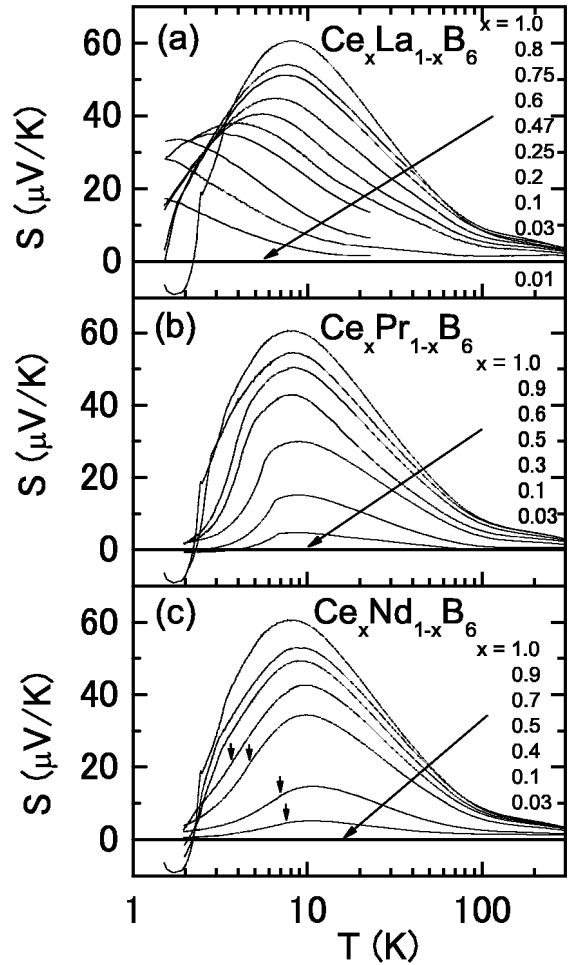

Fig. 1. Temperature dependence of $S$ of $Ce_xR_{1-x}B_6$ ($R=$ (a) La, (b) Pr and (c) Nd).

Here, we mention the overall characteristic behaviors of the $S$ of $Ce_xR_{1-x}B_6(R = La, Pr$ and Nd) in a paramagnetic region at high temperatures. In all the alloy systems, $S$ exhibits similar behaviors in which $S$ increases with decreasing temperature in a high-temperature region and the magnitude of $S$ is roughly proportional to the Ce concentration. $S(20 K)$ and $S(50 K)$ for three systems show a roughly inear decrease with a decrease in $x$ value. This is similar to the single-impurity character of the Kondo effect found in electrical resistivity, $\rho$.[21,24] Thus, the experimental results indicate that $S$ at high temperatures seems to be estimated as a summation of $S$ induced by the independent Kondo scattering on each Ce site.

With a decrease in Ce concentration, different behaviors are observed between $Ce_xLa_{1-x}B_6$ and $Ce_xR_{1-x}B_6(R = Pr, Nd)$ at low temperatures. In

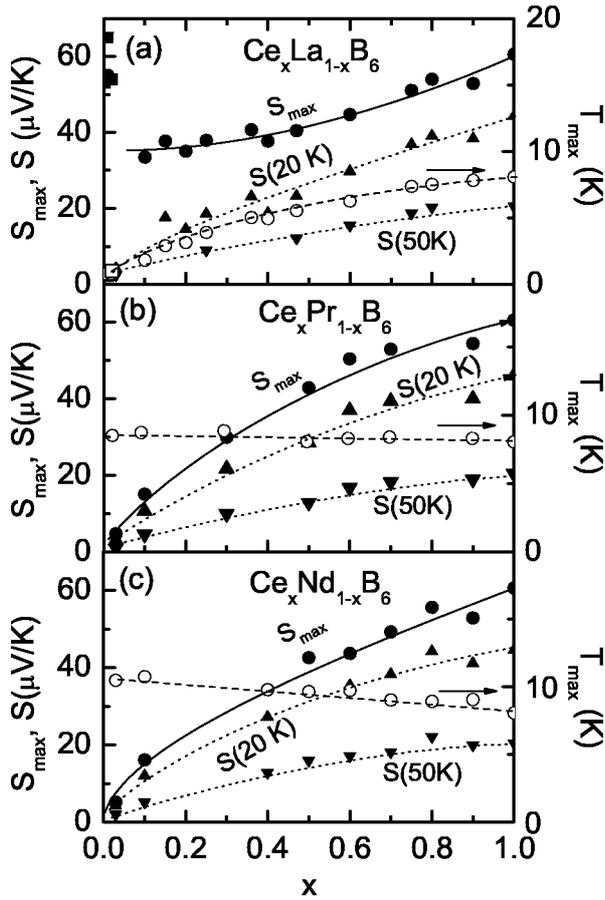

Fig. 2. $x$ dependences of $S_{\max}$(●), $S$ at $T$=20 K(▲) and 50 K(▼), and $T_{\max}$(○) for $Ce_xR_{1-x}B_6$. (a) $R$=La, (b) $R$=Pr and (c) $R$=Nd. All lines represent guides to the eyes.

$Ce_xLa_{1-x}B_6$, although $S_{\max}$ monotonically decreases down to $x \sim 0.1$, it does not converge to zero but increases largely in a very dilute $x$ region. $T_{\max}$ decreases from $\sim 8$ K for $x = 1.0$ to $\sim 2$ K for $x = 0.1$. For $x \leq 0.03$, $S_{\max}$ is not observed down to the lowest temperature studied in the present experiment. However, the $S$ values of these dilute samples are consistent with the previously reported results that a very large maximum in $S$ was found at around $T \sim 1$ K in more dilute compounds of $Ce_xLa_{1-x}B_6(x = 0.0025, 0.005, 0.01,$ and $0.02)$.[35] Filled and open squares for very dilute samples in Fig. 2(a) represent the $S_{\max}$ and $T_{\max}$ of the results from ref. 35. From the present and reported results, with a decrease in Ce concentration, the $T_{\max}$ of $Ce_xLa_{1-x}B_6$ seems to converge to $T_K \sim 1$ K in a very dilute region, which is expected from the theoretical study of $S$ for the Kondo compounds.[7,10] However, the results for $Ce_xR_{1-x}B_6(R = Pr, Nd)$ are very different from those for $Ce_xLa_{1-x}B_6$. The $S_{\max}$ of $Ce_xR_{1-x}B_6(R = Pr, Nd)$ decreases rather linearly with decreasing $x$ and converges to zero with a decrease in Ce concentration, and $T_{\max}$ shows a weak $x$ dependence both in $Ce_xPr_{1-x}B_6$ and $Ce_xNd_{1-x}B_6$, although it exhibits a small increase with decreasing $x$; from $\sim 8$ K to $\sim 9$ K in $Ce_xPr_{1-x}B_6$ and from $\sim 8$ K to $\sim 11$ K in $Ce_xNd_{1-x}B_6$. The $S$ of $Ce_xLa_{1-x}B_6$ in a small $x$ region at low temperatures is very different from those of $Ce_xR_{1-x}B_6$ ($R$ = Pr, Nd). This difference comes from the existence or nonexistence of the LRO phase transition at several Kelvins.

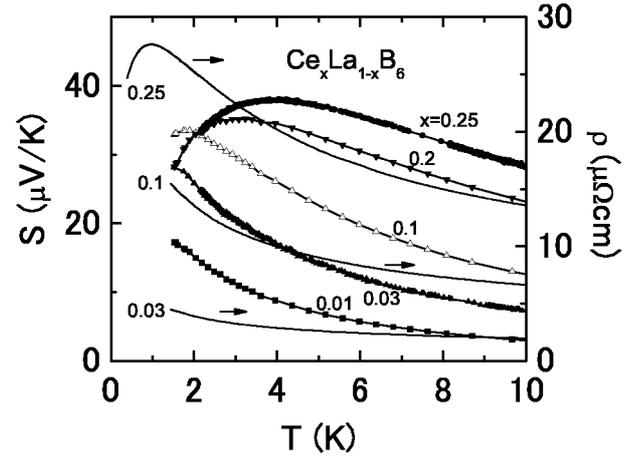

Fig. 3. Temperature dependences of $S$ (symbols) and $\rho$ (solid lines) for $Ce_xLa_{1-x}B_6$ ($x \leq 0.25$). The $\rho$ data for $x = 0.25$ is cited from ref. 22.

Figure 3 shows the $S$ and $\rho$ of $Ce_xLa_{1-x}B_6$ ($x \leq 0.25$) below 10 K. For $x = 0.25$, $\rho$ shows a maximum at $T \sim 1$ K. On the other hand, $S_{\max}$ locates at $T \sim 4$ K at which $\rho$ still exhibits a logarithmic increase with decreasing temperature. For $x = 0.1$, $S$ shows a maximum at $\sim 2$ K and a precursor of a maximum a slightly below 2 K for $x = 0.03$. In both samples, only a logarithmic increase is observed in $\rho$ in this temperature region. It should be noted that although the magnitude of $\rho$ is roughly proportional to the Ce concentration, that of $S$ does not exhibit such a proportional behavior in this temperature region.

Figure 4 shows the temperature dependences of the $\rho$ and $S$ of $LaB_6$. The residual resistivity $\rho_0$ of $LaB_6$ is as small as $\sim 0.2$ $\mu\Omega cm$, and $\rho$ shows an increase with increasing temperature above $\sim 50$ K. As for $S$, its magnitude is very small in the entire temperature region. A small temperature dependence is observed below $\sim 50$ K, but $S$ is almost constant above $\sim 50$ K.

In order to estimate the Ce contribution to $S$, usually the Nordheim-Gorter rule: $S \cdot \rho = S_{La} \cdot \rho_{La} + S_{Ce} \cdot \rho_{Ce}$, is used. In the present case, as $S_{La} \cdot \rho_{La}$ is very small in the entire temperature region, the observed $S$ in $Ce_xLa_{1-x}B_6$ could be estimated roughly as the contribution from the Ce ions.

3.2 $x$ dependences of $T_{\max}$ and long-range-ordering (LRO) temperatures

Figure 5 shows the $x$ dependences of $T_{\max}$ and the LRO temperatures, $T_{LRO}$, of $Ce_xR_{1-x}B_6$ ($R$ = La, Pr, Nd). $T_{\max}$ is shown by a dashed line in all the systems. The results on the LRO temperatures for $R$ = La and Nd are cited from refs. 22 and 24, respectively. The result for $R$ = Pr is obtained from the present result. Here, we note that the $x$ dependence of $T_{\max}$ is found to show roughly the same tendency as that of $T_{\mathrm{LRO}}$.

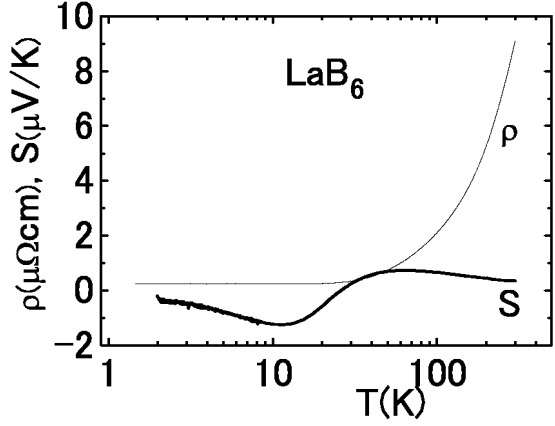

Fig. 4. Temperature dependences of $\rho$ (thin line) and $S$ (thick line) of $LaB_6$.

In $Ce_xLa_{1-x}B_6$, $T_Q$ is rapidly suppressed by La doping and disappears at around $x \sim 0.7$. The suppression rate of $T_N$ which is changed to $T^{\mathrm{IV-I}}$ at $x \sim 0.8$ is smaller than that of $T_Q$ and seems to disappear at around $x \sim 0.3$.[22] However, even below $x \sim 0.3$, the effect of the Ce-Ce interaction remains, as is seen in a broad maximum of $\rho$ at $T \sim 1$ K for $x = 0.25$ in Fig. 3.

In $Ce_xPr_{1-x}B_6$, for $0.88 \leq x \leq 1$, $T_Q$ decreases, but $T_N$ increases with decreasing $x$. With a further decrease in Ce concentration below $x \leq 0.88$, two successive magnetic phase transitions, $T_{\mathrm{IC1}}$ and $T_{\mathrm{IC2}}$, appear, and both transition temperatures increase monotonically with decreasing $x$. The $T_{\mathrm{Cm}}$ of $PrB_6$ is rapidly suppressed by a small amount of Ce doping. Recently, we have reported the magnetic phase diagram of $Ce_{0.7}Pr_{0.3}B_6$ and showed that the quadrupolar ordering coexists with the magnetic ordering.[23] This indicates that both Ce and Pr ions contribute to the formation of the LRO states and the Ce-Pr coupling in $Ce_xPr_{1-x}B_6$ differs strongly from the weak Ce-Nd coupling in $Ce_xNd_{1-x}B_6$ as will be discussed later.

In $Ce_xNd_{1-x}B_6$,[24] for $0.88 \leq x \leq 1$, the $x$ dependences of $T_Q$ and $T_N$ are similar to those of $Ce_xPr_{1-x}B_6$. With a further decrease in Ce concentration, two magnetic LRO phases appear below $T_N^H$ and $T_N^L$ for $0.3 \leq x \leq 0.88$. $T_N^L$ shows a decrease with decreasing $x$ and seems to disappear at $x \sim 0.3$, which is the same value at which phase IV disappears in $Ce_xLa_{1-x}B_6$.

### 3.3 Low-temperature behavior of $S$ of $Ce_xR_{1-x}B_6$

In order to observe the evolution of the Kondo effect and LRO with the change of Ce concentration in $Ce_xR_{1-x}B_6(R = Pr$ and Nd$)$, we show the temperature dependence of $S$ at low temperatures in an expanded scale in Figs. 6 and 7, respectively.

First, we briefly mention the results for $Ce_xLa_{1-x}B_6$. A decrease in $S$ is observed below $T_{\mathrm{LRO}}$ for $x \geq 0.75$. Although such a decrease is not observed for $x \leq 0.6$, it is because their $T_{\mathrm{LRO}}$ is lower than the lowest temperature studied in the present experiment.

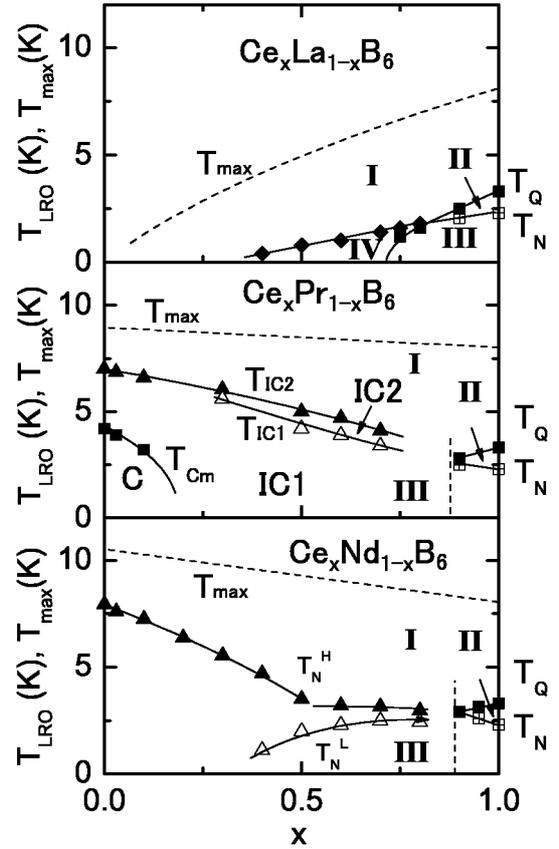

Fig. 5. $x$ dependences of $T_{\max}$ (dashed line) and LRO temperatures (symbols) of $Ce_xR_{1-x}B_6(R = La$, Pr and Nd$)$ at $H=0$. The LRO temperatures for $R = La$ and Nd are cited from refs. 22 and 24, respectively.

Next, we mention the results for $Ce_xPr_{1-x}B_6$. For $x = 0.9$, only one anomaly is found at $\sim 2.8$ K because $T_Q$ is very close to $T_N$ in this sample. For $0.3 \leq x \leq 0.7$, $S$ shows a decrease below $T_{\mathrm{IC1}}$ and $T_{\mathrm{IC2}}$. The decrease in $S$ below $T_{\mathrm{IC1}}$ is much more pronounced than that below $T_{\mathrm{IC2}}$. A clear decrease below $T_{\mathrm{IC1}}$ and $T_{\mathrm{IC2}}$ indicates that Ce ions lose their character as the Kondo scattering center and contribute to the IC1 or IC2 magnetic ordering as localized ions. The concave curvature below $T_{\mathrm{IC1}}$ or $T_{\mathrm{IC}}$ becomes more pronounced below $x = 0.3$, which is very different from no anomaly at $T_N^H$ in $Ce_xNd_{1-x}B_6(x \leq 0.5)$.

For $Ce_xNd_{1-x}B_6$, with a decrease in Ce concentration from $x = 1$ down to 0.9, the sharp drop at $T_N$ moves up to a slightly higher temperature, while the anomaly at $T_Q$ oppositely moves down to lower temperatures. As $x$ is decreased through $x \sim 0.88$, new LRO phases appear at $T_N^L$ and $T_N^H$.[25] For $x = 0.6$ and 0.7, the LROs at $T_N^L$ and $T_N^H$ exhibit sharp and weak changes in slope, respectively. These behaviors of $S$ at $T_N^L$ and $T_N^H$ are consistent with the results on $\rho$, which exhibits a sudden drop at $T_N^L$ and a lax decrease at $T_N^H$.[24] On the other hand, $S$ for $x=0.5$ shows a clear anomaly at $T_N^L$ but no anomary at $T_N^H$. These behaviors at $T_N^L$ and $T_N^H$ are consistent with the results on $\rho$, which shows a clear decrease at $T_N^L$ but no anomaly at $T_N^H$. Thus, the present results also suggest that the nature of the LRO state below $T_N^H$ is changed below and above $x \sim 0.5$. It is noted that the magnetic susceptibility and specific heat show clear anomalies at $T_N^H$ for $x=0.5$.

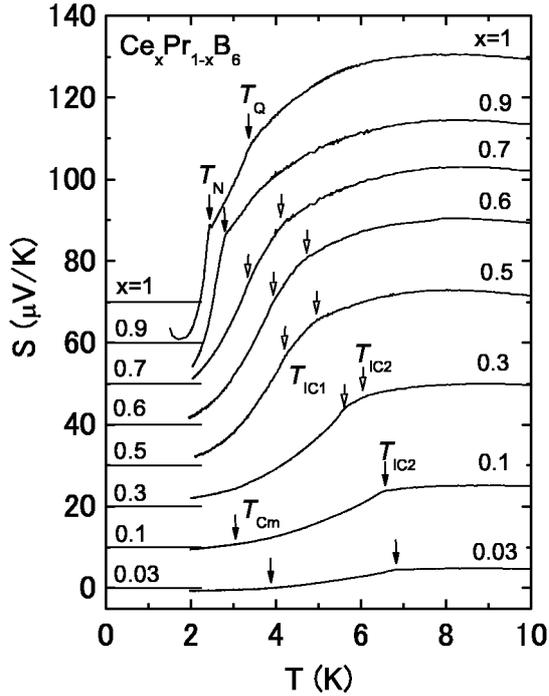

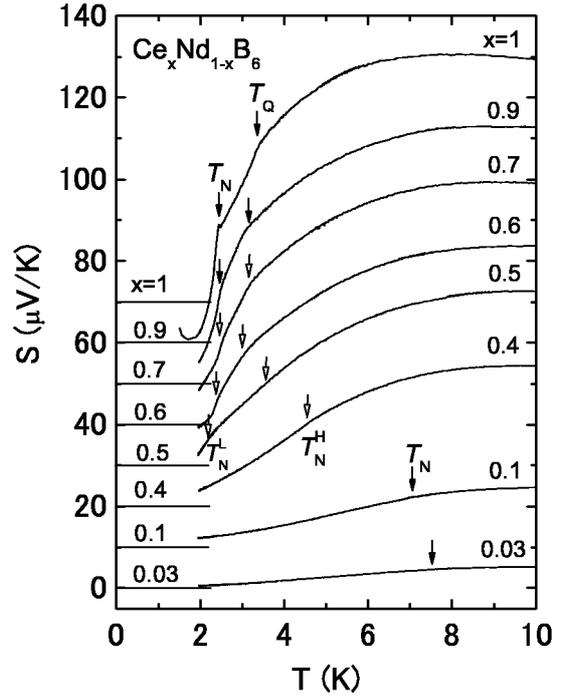

Fig. 6. Temperature dependence of $S$ below $T=10$ K for $Ce_xPr_{1-x}B_6$. For clarity, the origin of the vertical axis is shifted. Each horizontal line indicates the origin of the vertical axis for each compound. Arrows indicate the LRO transition temperatures.

Fig. 7. Temperature dependence of $S$ below $T=10$ K for $Ce_xNd_{1-x}B_6$. For clarity, the origin of the vertical axis is shifted. Each horizontal line indicates the origin of the vertical axis for each compound. Arrows indicate the LRO transition temperatures, which are cited from ref. 24.

4. Discussion

In the high-temperature region, the magnitude of $S$ originating from the Kondo effect is roughly proportional to the Ce concentration in all the alloy systems. This suggests that the impurity Kondo character is conserved and the Ce-Ce or Ce-$R$($R$ = Pr, Nd) interaction is negligible at high temperatures.

On the other hand, at low temperatures, very different $x$ dependences of $S_{max}$ and $T_{max}$ are observed between $Ce_xLa_{1-x}B_6$ and $Ce_xR_{1-x}B_6$($R$ = Pr, Nd). The maximum of $S$ is often used to estimate the $T_K$ of a dilute Kondo system. It is reported that the dilute compounds $Ce_xLa_{1-x}B_6$($x = 0.0025 \sim 0.02$) exhibit a large peak slightly below $T_K$,[35] consistent with the theoretical study.[7] However, in $CeB_6$, $T_{max}$ is observed at $\sim 8$ K, which is much higher than the $T_K$ of the dilute $Ce_xLa_{1-x}B_6$ alloys. The present results show that $T_{max}$ exhibits a large variation as a function of $x$. On the other hand, it was reported that $T_K \sim 1$ K is almost constant independent of $x$ in $Ce_xLa_{1-x}B_6$.[21] However, strictly speaking, it has not been necessarily verified.

As the possible origin of $S_{max}$, the entrance into the coherent Kondo state is considered. However, this possibility is ruled out by comparing the results of $Ce_xLa_{1-x}B_6$ with those of $Ce_xR_{1-x}B_6$($R$ = Pr, Nd). In $Ce_xR_{1-x}B_6$, being different from $Ce_xLa_{1-x}B_6$, $T_{max}$ exhibits a small increase from 8 K to $\sim 10$ K with decreasing $x$. In $Ce_xR_{1-x}B_6$($R$ = Pr, Nd) where Ce ions are surrounded by the magnetic ions, it is difficult to expect that the entrance into the coherent Kondo state is conserved at the same temperature as $\sim 8$ K.

Next, we discuss the possibility that $S_{max}$ appears at $T_K$, namely, the possibility that $T_{max} \sim T_K$ in $Ce_xR_{1-x}B_6$. Generally, $T_K$ is varied with the change of the effect of chemical pressure in the alloy systems. The effect of chemical pressure on the Kondo effect is important for discussing the physical properties of the alloy systems. Such a chemical pressure effect was discussed in $Ce_x(La, Y)_{1-x}Cu_{2.05}Si_2$ in detail.[14,15] It is known that $T_K$ is enhanced under pressure, which is a result of the enhancement of the mixing between the $4f$ and conduction electrons. In the alloy systems, the difference of the ionic radius of the doped rare-earth ion from that of the mother compound induces the chemical pressure effect. When $x$ decreases, the negative pressure effect is expected in $Ce_xLa_{1-x}B_6$ and the positive one in $Ce_xR_{1-x}B_6$($R$ = Pr, Nd). On the other hand, in a large $x$ region, the effect of pressure on the Ce-Ce interaction becomes larger in addition to the single-ion pressure effect. The effect of pressure on the Ce-Ce interaction consists of two different mechanisms. One is to form the coherent Kondo state and the other is to increase the Ce-Ce interaction as a localized system which suppresses the Kondo effect.

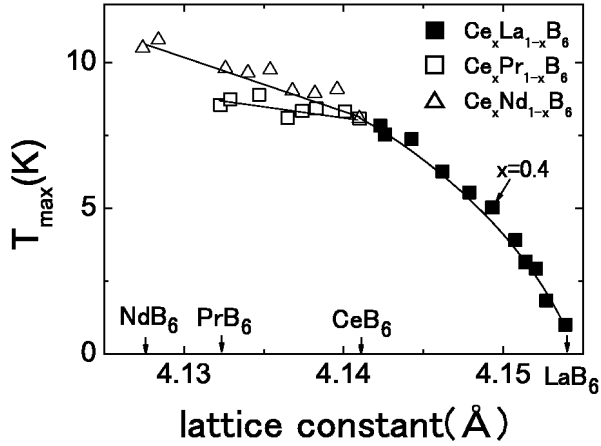

Fig. 8. Lattice constant dependence of $T_{\max}$ of $Ce_xR_{1-x}B_6$ ($R$ = La(■), Pr(□) and Nd(△)).

Experimental results are as follows. $T_{\max}$ decreases in $Ce_xLa_{1-x}B_6$ but slightly increases in $Ce_xR_{1-x}B_6$ ($R$ = Pr, Nd) with decreasing $x$. This tendency could be explained by the consideration that $T_K$ is enhanced when the average lattice constant decreases. However, such a consideration seems to be unacceptable, when we consider the relationship between $T_{\max}$ and the lattice constants of $Ce_xR_{1-x}B_6$ as follows. Here, we show the lattice constant dependence of $T_{\max}$ in Fig. 8. $T_{\max}$ shows a large increase with a decrease in lattice constant in a small $x$ region of $Ce_xLa_{1-x}B_6$ and its rate of increase becomes smaller as $x$ approaches 1. The slope of $T_{\max}$ vs lattice constant seems to change at around $x \sim 0.4$. This suggests that the slope of $T_{\max}$ becomes small when the LRO state appears at low temperatures. The $T_{\max}$ of $Ce_xR_{1-x}B_6$ increases with a decrease in lattice constant, but the slope of $Ce_xR_{1-x}B_6$ is smaller than that of $Ce_xLa_{1-x}B_6$. The slope of $T_{\max}$ differs between $Ce_xPr_{1-x}B_6$ and $Ce_xNd_{1-x}B_6$. The $T_{\max}$ values of these two systems are not scaled by the lattice constant. That for $Ce_xPr_{1-x}B_6$ is slightly smaller than that for $Ce_xNd_{1-x}B_6$. Thus, it is difficult to consider that $T_{\max} \sim T_K$.

We expect another mechanism for the $x$ dependence of $T_{\max}$, which may originate from the difference of Pr and Nd ions of these two systems. Here, we note that the $x$ dependence of $T_{\max}$ shows roughly the same tendency as that of $T_{LRO}$ as shown in Fig. 5. In $Ce_xLa_{1-x}B_6$, $T_{\max}$ decreases with decreasing $x$. The Ce-Ce interaction is suppressed with decreasing $x$ because of the increase in the Ce-Ce distance. It is possible to associate $T_{\max}$ with the temperature at which the effect of the Ce-Ce interaction begins to appear in the macroscopic properties. In $CeB_6$, below $\sim 10$ K, $\rho$ exhibits a deviation from the ln $T$ dependence at high temperatures.[21] This may be due to the suppression of the Kondo effect by the Ce-Ce interaction. This temperature of $\sim 10$ K roughly coincides with $T_{\max}$ in $CeB_6$. A similar relationship between the $T_{\max}$ and the temperature at which the deviation of $\rho$ from the ln $T$ dependence in the high-temperature region is also observed in $Ce_xLa_{1-x}B_6$ in Fig. 1(b) of ref. 21. With decreasing Ce concentration in $Ce_xLa_{1-x}B_6$, the Ce-Ce interaction is suppressed, which may suppress $T_{\max}$. Thus, the appearance of $S_{\max}$ may be a result of the competition between the Kondo effect and the Ce-Ce interaction. However, in $Ce_xR_{1-x}B_6$ ($R$ = Pr, Nd), $T_{\max}$ shows a small increase with decreasing $x$, different from the roughly linear $x$ dependence in $Ce_xLa_{1-x}B_6$. The clear difference between $Ce_xLa_{1-x}B_6$ and $Ce_xR_{1-x}B_6$ ($R$ = Pr, Nd) is the nonexistence or existence of the magnetic ions between Ce ions. When the interaction between the rare-earth ions is that of the RKKY type, we expect the suppression of the Ce-Ce interaction in $Ce_xLa_{1-x}B_6$ with decreasing Ce concentration because of the increase in Ce-Ce distance. On the other hand, in $Ce_xR_{1-x}B_6$ ($R$ = Pr, Nd), $R$ ions exist between Ce ions. The experimental result that $T_{\max}$ does not depend so much on $x$ in $Ce_xR_{1-x}B_6$ ($R$ = Pr, Nd) suggests that the strength of the Ce-Ce interaction is not markedly reduced by the exsitence of the magnetic $R$ ions.

As for the results for $Ce_xLa_{1-x}B_6$ for $0.1 \leq x \leq 0.25$, there exists another question why $T_{\max}$ is much higher than the temperature at which $\rho$ exhibits a maximum, which was shown in Fig. 3. The maximum of $\rho$ may originate from the growth of the Ce-Ce interaction as mentioned above. It is also possible to associate $S_{\max}$ with the growth of the Ce-Ce interaction. These suggest that the effect of the growth of the Ce-Ce interaction is more sensitive for $S$ than for $\rho$.

Finally, we discuss $S$ in the LRO state. In the case of $Ce_xLa_{1-x}B_6$, a clear decrease is observed in $S$ below $T_Q$, $T_N$ or $T^{IV-I}$. Such a decrease originates from the destruction of the Kondo singlet state accompanied by the LRO induced by Ce ions themselves. In $Ce_xPr_{1-x}B_6$, $S$ shows a clear decrease below $T_{LRO}$. This suggests that the Kondo singlet state on the Ce site is destroyed below $T_{LRO}$ and both Ce and Pr ions equally contribute to the formation of the LRO state. This indicates that the Ce-Pr coupling is strong in $Ce_xPr_{1-x}B_6$. On the other hand, in $Ce_xNd_{1-x}B_6$, although a similar anomaly of $S$ accompanied by the LRO is observed for $x \geq 0.6$, an anomaly is difficult to recognize at $T_N^H$ for $x \leq 0.5$. This indicates that the electronic state of the Ce ion below $T_N^H$ is different below and above $x \sim 0.5$, where the $x$ dependence of $T_N^H$ is changed as is shown in Fig. 5. For $x \geq 0.6$, the Kondo singlet state is destroyed below $T_N^H$. For $x \leq 0.5$, the Kondo state survives below $T_N^H$ and is destroyed below $T_N^L$. In our previous papers,[24,25] we reported that the LRO state below $T_N^L$ is formed by Ce ions themselves but that below $T_N^H$ is mainly formed by Nd ions and we discussed that the Kondo effect survives between $T_N^L$ and $T_N^H$. The present results support these considerations.

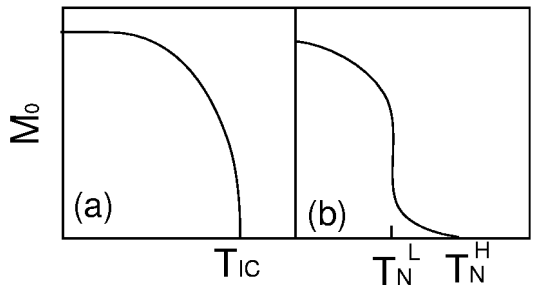

Fig. 9. Conjectured temperature dependence of magnetic ordered moment on Ce site in (a) $Ce_xPr_{1-x}B_6$ and (b) $Ce_xNd_{1-x}B_6$ with $0.3 \leq x \leq 0.5$ below LRO temperatures. See the text for details.

From these results, we conjecture the temperature dependence of the ordered moment on the Ce site below the $T_{LRO}$ of $Ce_xR_{1-x}B_6 (R = Pr, Nd)$ as shown in Figs. 9(a) and 9(b), respectively. In $Ce_xPr_{1-x}B_6$, as both Ce and Pr ions equally contribute to the LRO, the ordered moment on the Ce site shows a normal behavior as shown in Fig. 9(a). However, in $Ce_xNd_{1-x}B_6 (0.3 \leq x \leq 0.5)$, the effect of the ordered Nd ions on Ce ions is small and the Kondo singlet state on the Ce site survives below $T_N^H$. As the Kondo singlet state is expected to be destroyed at very low temperatures, the temperature dependence of the ordered moment on the Ce site is expected as that shown in Fig. 9(b).

At present, we do not know why such a large difference in behavior is observed between $Ce_xPr_{1-x}B_6$ and $Ce_xNd_{1-x}B_6$ below $T_{LRO}$.

## 5. Conclusion

In summary, we have measured the thermopower, $S$, of $Ce_xR_{1-x}B_6 (R = La, Pr, Nd)$ in a wide range of the Ce concentration. At high temperatures, the impurity Kondo behavior is observed in the $S$ of all the alloy systems, as was observed in the electrical resistivity of $Ce_xLa_{1-x}B_6$. At low temperatures, $S$ exhibits a maximum at $T_{max}$ in all the alloy systems, but their temperature dependences differ from each other. In $Ce_xLa_{1-x}B_6$, both $S_{max}$ and $T_{max}$ decrease with decreasing $x$, but $S_{max}$ increases below $x \sim 0.1$. $T_{max}$ seems to converge to the $T_K \sim 1$ K of very dilute $Ce_xLa_{1-x}B_6$. In $Ce_xR_{1-x}B_6 (R = Pr, Nd)$, $T_{max}$ shows a weak $x$ dependence but $S_{max}$ seems to converge to zero in a sample with very small $x$ value. The $x$ dependnce of $T_{max}$ is consistent with the chemical pressure effect in the alloy systems but is not explained only by this effect. As another origin of the $x$ dependence of $T_{max}$, we proposed the effect of the Ce-Ce interaction. In $Ce_xLa_{1-x}B_6$, the Ce-Ce interaction is suppressed with decreasing $x$, but in $Ce_xR_{1-x}B_6 (R = Pr, Nd)$, it is conserved by the existence of the $R$ ions between Ce ions. $T_{max}$ is higher than the temperature at which $\rho$ exhibits a maximum for $0.1 \leq x \leq 0.25$, which suggests that the effect of the Ce-Ce interaction for $S$ appears at higher temperatures than for $\rho$. In the LRO state, very different behaviors are observed between $Ce_xPr_{1-x}B_6$ and $Ce_xNd_{1-x}B_6$. In $Ce_xPr_{1-x}B_6$, $S$ is rapidly suppressed below the LRO temperature, indicating the suppression of the Kondo singlet state and the strong Ce-Pr coupling. On the other hand, in $Ce_xNd_{1-x}B_6 (x \leq 0.5)$, $S$ is not affected by the magnetic ordering mainly induced by Nd ions, indicating the survival of the Kondo singlet state and the weak Ce-Nd interaction in this system.

## Acknowledgments

The present work was supported by a COE Research (13E2002) from the Ministry of Education, Culture, Sports, Science and Technology of Japan. MSK was financially supported by the Japan Society for the Promotion of Science. The low-temperature measurements were supported by the Cryogenic Center of Hiroshima University.